\documentclass{article}
\newcommand{\bfr}{\begin{flushright}}
\newcommand{\efr}{\end{flushright}}
 
\begin{document}
\title{Phase Transition and String Formation in Six-Dimensional Gauge
Theory
}
\author{ 
Atsushi Nakamula\\
Department of Physics, Tokyo Metropolitan University,\\
Setagaya-ku, Tokyo 158, Japan\\
and\\
Kiyoshi Shiraishi\\
Institute for Nuclear Study, University of Tokyo, \\
Midori-cho, Tanashi,
Tokyo 188, Japan
}
\date{Prog. Theor. Phys. {\bf 84} (1990) pp. 1100--1107
}
\maketitle
\begin{abstract}
We consider an $SU(2)$ gauge theory in six-dimensions. We show that there
exists non-trivial structure of gauge vacua in compactified background
$M_4\times S^2$. In this circumstances, there can exist a cosmic string,
whose solution is recently constructed by the present authors. We
consider the stage of the formation of the strings, i.e., the phase
transition of the gauge configuration on the sphere in the `hot' early
universe. We find that if $SU(2)$-doublet fermions are the dominant
component of the radiation at high temperatures, the phase transition
successively occurs after the spontaneous-compactification transition.
Otherwise, non-trivial vacuum with `monopole' configuration of the
$SU(2)$ gauge field never appears globally and strings must be formed at
the same time as the extra space is compactified.
\end{abstract}

\bigskip

The idea of the existence of extra dimensions in particle physics has
a long history since 1919.\cite{1,2} In recent years, the development of
string theories revealed the necessity of multi-dimensional space-time
for a simple formulation of the dynamics of the string.\cite{3} If the
string theory, as a unified theory of the forces, describes the
physics of the real world, then the extra dimensions other than four
must be hidden from our observations; we usually consider them to be
curled up to a tiny scale. We call the phenomenon as
``compactification''.\cite{2,4}

In higher-dimensional theories, there is generic problems of fermions.
Compactifications in curved spaces in general lead to huge masses for fermions.
Moreover, it is difficult to realize the chirality of femions.\cite{5} To avoid the problem,
we must introduce elementary gauge fields and non-trivial configurations
of them on the compactified space. Elementary gauge fields are certainly
contained in the effective fleld theory of the string theory of the
so-called realistic type, such as the heterotic string theory.\cite{6}
 
In several models with compactification, gauge fields play a role in preventing the
extra space from collapsing.\cite{7} In models of other type, non-trivial topology of the
extra space admits a vacuum configuration of the gauge field to break its own
symmetry.\cite{8}

We consider in this paper a non-trivial configuration the gauge field in the
compactified extra space.\cite{9,10} This leads to a formation of topological defects in
our space. Presumably this will give a new insight on the interplay of the topology of the
gauge fields and the structure of the space.
 
We work with $SU(2)$ gauge symmetry as the simplest non-Abelian group. One
may consider this $SU(2)$ group as a subgroup of a larger gauge group. In this paper,
we consider six-dimensional space-time. Among six dimensions, two
spatial dlmensions are assumed to be compactiied into the sphere.
 
The radius of the compactifled space $S^2$ is a constant, denoted as
$b$, if the gravitational interaction is neglected. One can suppose a
non-trivial gauge configuration on $S^2$, besides the trivial one. ]t is a
``monopole'' configuration:\cite{9}
\begin{equation}
{\bf A}_\phi=\frac{1}{2}
\left(
\begin{array}{cc}
1-\cos\theta & 0\\
0 & -(1-\cos\theta) 
\end{array}
\right)\,,  
\label{eq1}
\end{equation}
where $\theta$ and $\phi$ are the polar and the azimuthal angles of the internal sphere.
These coordinates are denoted by indices $m$ and $n$. This configuration has considerable
symmetry in itself. For instance, fermion fields have zero modes in general if the
fields couple to the background field (\ref{eq1}) and the metric.\cite{9,7} The gauge
symmetry is, however, reduced to $U(1)$ symmetry of which gauge field is expressed as the
following form:
\begin{equation}
{\bf A}_\mu=A_\mu(x^\mu)\frac{1}{2}
\left(
\begin{array}{cc}
1 & 0\\
0 & -1 
\end{array}
\right)\,,  
\label{eq2}
\end{equation}
where $A_\mu(x^\mu)$ depends only on the coordinates of four dimensions but not on the
extra dimensions. Subscript $\mu$ runs over the four coordinates.

The vacuum associated with the configuration (\ref{eq1}) has higher energy density than
the trivial gauge vacuum ${\bf A}_m=0$, because of $\langle {\rm Tr~}{\bf
F}_{mn}{\bf F}^{mn}\rangle=1/b^4$, where the field strength ${\bf F}_{mn}=\partial_m{\bf
A}_n-\partial_n{\bf A}_m+i[{\bf A}_m, {\bf A}_n]$.
 
Next We consider a global view of the vacua of the $SU(2)$ gauge configurations,
i.e., we investigate the path attached to both the trivial and the non-trivial gauge
vacua.

We make an ansatz for the gauge fields and parametrize them in the
matrix form as \cite{10}
\begin{eqnarray}
{\bf A}_\theta&=&\frac{1}{2}
\left(
\begin{array}{cc}
0 & -i\Phi e^{-i\phi}\\
i\Phi^* e^{i\phi} & 0 
\end{array}
\right)\,,  
\label{eq3a}\\
{\bf A}_\phi&=&-\frac{1}{2}
\left(
\begin{array}{cc}
0 & \Phi e^{-i\phi}\\
\Phi^* e^{i\phi} & 0 
\end{array}
\right)\sin\theta+\frac{1}{2}
\left(
\begin{array}{cc}
1-\cos\theta & 0\\
0 & -(1-\cos\theta) 
\end{array}
\right)\,,  
\label{eq3b}\\
{\bf A}_\mu&=&A_\mu(x^\mu)\frac{1}{2}
\left(
\begin{array}{cc}
1 & 0\\
0 & -1 
\end{array}
\right)\,,  
\label{eq3c}
\end{eqnarray}
where $\Phi$ is a complex variable and does not depend on the coordinates of the extra
dlmensions.

Here $\Phi$ connects the monopole vacuum with the trivial vacuum. It is obvious
that the monopole configuration realizes when $\Phi=0$. When $|\Phi|=1$, the configuration
is equivalent to the trivial vacuum ${\bf A}_m=0$ modulo gauge transformations. For
example, for $\Phi=1$, we find
\begin{equation}
{\bf A}'_m=\Omega{\bf A}_m\Omega^\dagger-i\Omega\partial_m\Omega^\dagger=0
\label{eq4a}
\end{equation}
with
\begin{equation}
\Omega=
\left(
\begin{array}{cc}
\cos(\theta/2) & -\sin(\theta/2) e^{-i\phi}\\
\sin(\theta/2) e^{i\phi} & \cos(\theta/2) 
\end{array}
\right)\,.
\label{eq4b}
\end{equation}

Now we consider that $\Phi$ depends on the coordinates of the four dimensions. The
above parametrization reduces the Yang-Mills equation to
\begin{eqnarray}
& & D^\mu D_\mu\Phi+\frac{1}{b^2}(1-|\Phi|^2)\Phi=0\,,\label{eq5a}\\
& & \partial_\mu(b^2 F^{\mu\nu})+i(\Phi^*D^\nu\Phi-\Phi D^\nu\Phi^*)=0\,,                
\label{eq5b}
\end{eqnarray}
where $F_{\mu\nu}=\partial_\mu A_\nu-\partial_\nu A_\mu$. The covariant derivative is
defined as $D_\mu=\partial_\mu+iA_\mu$.
 
we find that both $\Phi=0$0 and $|\Phi|=1$ are the stationary points of the effective
``Higgs'' potential, $V(\Phi)\propto (1-|\Phi|^2)^2$. The point $\Phi=0$ corresponds to a
local maximum of the potential.
 
These equations (\ref{eq5a}, \ref{eq5b}) are the same as those in the Abelian-Higgs
model \cite{11} except for the radius of the compact space. At this stage, the original
$SU(2)$ symmetry reduces to $U(1)$ gauge symmetry, if $\langle\Phi\rangle=0$.
 
Note again that the vacuum associated with $\langle|\Phi|\rangle=1$ is gauge equivalent to
the trivial vacuum, i.e., $\langle{\bf A}_m\rangle=0$ in the original notation of the
$SU(2)$ gauge field. Thus all the $SU(2)$ gauge fields remain massless. This seems
contradictory, since the effective $U(1)$ gauge fleld $A_\mu$ in (\ref{eq5a}, \ref{eq5b}),
which comes from the original $SU(2)$ gauge field, undoubtedly becomes massive due to the
coupling to $\Phi$. In the present context, however, it can be understood as
an example of the well-known mechanism that massive Kaluza-Klein
modes in the vacuum with $\Phi=0$ become massless when the ``Higgs'' takes
the value $|\Phi|=1$. A simple example of the mechanism can be found in the
model with torus compactification, and there the vacuum gauge field on
the torus shifts the Kaluza-Klein modes.\cite{18}
 
Nevertheless, we emphasize the transition for two reasons. First, the
formation of strings can be asssociated with this stage. We can
construct a vortex solution in the effective Abelian-Higgs system governed
by Eqs.~(\ref{eq5a}, \ref{eq5b}).\cite{10} If the strings play some roles in the early
universe, the investigation of their formation is important. Second, fermion
fields coupled to the gauge field has zero-mass spectrum in the
``symmetric'' vacuum with $\Phi=0$.\cite{9,7} If there is no $SU(2)$ gauge
background, fermions acquire their mass because of the curvature of
the extra space.\cite{5} Then, our parametrization allows the
interpretation that all the fermion mass is generated from the
vacuum expectation value of the ``higgs'' field $\Phi$. This interpretation
is useful if we compare fermion zero-modes around our cosmic strhg
with those of a usual type.
 
Let us examine what the vortex configuration looks like. For a while we neglect
the coupling to gravity. We consider the vortex-like configuratiun which lies along
the $z$-axis of a cylindrical coordinate system $(r, \psi, z)$. We take the rotationally
symmetric ansatz for the $n$-vortex configuration,
\begin{eqnarray}
\Phi&=&X(r) e^{in\psi}\,,\label{eq6a}\\
A_\psi(r)&=&n(P(r)-1)\,,\quad\mbox{otherwise  } A_\mu(r)=0\,.
\label{eq6b}
\end{eqnarray}

The asymptotic values for the ``scalar'' and ``vector'' fields are
\begin{equation}
X(0)=0\,,\quad P(0)=1
\label{eq7a}
\end{equation}
and
\begin{equation}
X(\infty)=1\,,\quad  P(\infty)=0\,. 
\label{eq7b}
\end{equation}
 
The solution of Eqs.~(\ref{eq5a}, \ref{eq5b}) with these boundary conditions is the
Nielsen-Olesen vortex \cite{11,12}, in the Abelian-Higgs model with the implicit critical
relation between gauge and Higgs couplings.
 
As mentioned above, we wish to emphasize that, apart from the Kaluza-Klein
gauge symmetry, the original $SU(2)$ symmetry is broken to $U(1)$ at the core of the
string, while the $SU(2)$ is unbroken at radial infinity.
 
Note that an $SU(2)$ gauge field forms the non-trivial configuration in a conformally flat
four-space, which is called ``(anti-) instanton''.\cite{12,13} We construct a
topologically non-trivial configuration of the $SU(2)$ gauge field in the four-space which
is composed of $x$-$y$ plane and the compactified sphere. Therefore, it can be said that the
string is made from a deformed instanton. Actually it is easy to see that the topological
winding number is the same as that of the (anti-)instanton.\cite{14} This is closely
connected with the number of fermionic zeto modes along with the string. The
fermion zero modes can affect the property of the string. This subject is reported
elsewhere.\cite{14}

If we want to couple gravity to the string, we must introduce a cosmological
constant and a $U(1)$ gauge field, for example, in order to guarantee the asymptotic
form of the space-time as (flat four dimensional space-time) $\times S^2$. When the ratio of
the $SU(2)$ and $U(1)$ couplings is finite, $b$ is no longer a constant and the metric of
the space-time becomes deformed accoding to the Einstein equations. Equations (\ref{eq5a})
and (\ref{eq5b}) differ from those in the Abelian-Higgs model because of the non-trivial
$r$-dependence of $b$. The numerical solution of these equations was previously reported
by the present authors.\cite{10}
 
Now, we consider the condition of the phase transitions and the effect of fermions
as matter fields.
 
There are two possible phase transitions in our model. One is the spontaneous
compactification and the other is the gauge symmetry breaking. To put
our model in a cosmological context, We assume that the universe was
very hot at its initial stage. Under the assumption, we consider
quantum effects of mattcr fields at high temperature as usual in
grand unified theories.\cite{15}
 
First we consider compactification. It is known that at high temperature, the
pressure of matter fields can destabilize the compactification of the
extra dimensions.\cite{16} Although the decisive mechanism of
compactification in the early history of the universe is unknown, We
may consider the possibility that quantum tunneling took place in hot,
multidimensional universe.\cite{17} In this scenario, the transition to
compactified space occurs when the temperature decreases below a
critical temperature $T_{comp}$. Let us compute the value of $T_{comp}$. From
now on, as in the analysis of phase transitions in usual field
theories, we adopt the approximation that all back-ground values of the
fields are spatially homogeneous.
 
In the following analysis, We introduce $N_F$ fermion doublets as matter fields.
This situation corresponds to the dominance of the fermionic matter over the bosonic
contributions. We follow the analysis of Ref.~\cite{6}. We calculate the
free energy density by using high-temperature expansion.
 
The square of the line element is set to be
\begin{equation}
ds^2= -dt^2+a^2(t)d\Omega^2+ b^2(t)(d\theta^2+\sin^2\theta d\phi^2)\,,
\label{eq8}
\end{equation}
where $d\Omega^2$ represents the line element of the isotropic and homogeneous three-space.
Here $a$ and $b$ are the scale factors of the noncompact and the compact space,
respectively.
  
We take assumptions that the scale factor of our space is much larger than that
of the extra space, i.e., $a\gg b$. It is also assumed that the metric is quasi-static,
i.e., $(\dot{b}/b)^2, (\dot{a}/a)^2\ll 1/b^2$.

In the calculation of the free energy, we must note that when $\Phi=0$ $SU(2)$ gauge
configuration is magnetic monopole on $S^2$. After simple calculation we
get the expressions of the density of the free energy at temperature $T$
for each gauge background,\cite{18}
\begin{eqnarray}
f(|\Phi|=1)&=&-N_F\left(\frac{31\pi^3}{945}T^6-\frac{7\pi}{1080}\frac{1}{b^2}T^4+\cdots
\right)\,,\label{eq9a}
\\ f(\Phi=0)&=&-N_F\left(\frac{31\pi^3}{945}T^6+\frac{7\pi}{540}\frac{1}{b^2}T^4+\cdots
\right)\,,     
\label{eq9b}
\end{eqnarray}
where we exhibit the high temperature expansion up to the next-to-leading terms for
later use. In obtaining $T_{comp}$, we use the leading terms.
 
To accomplish the compactification at zero temperature, we employ the
following action including the Einstein gravity plus $U(1)$ gauge
field:\cite{10}
\begin{eqnarray}
S&=&\int d^6x\sqrt{-g}\left(-\frac{1}{2\kappa^2}R+\frac{1}{4e^2}{\rm
Tr~}({\bf F}_{MN}{\bf F}^{MN})+\frac{1}{4g^2}G_{MN}G^{MN}+\lambda\right)\nonumber \\
& &+S(\mbox{matter fields})\,.
\label{eq10}
\end{eqnarray}
Here $F_{MN}$ and $G_{MN}$ are the field strengths of $SU(2)$ and$U(1)$ gauge fields,
respectively. $e$ and $g$ are the gauge coupling constant of $SU(2)$ and $U(1)$,
respectively; $\lambda$ is a cosmological constant. We assume the $U(1)$ gauge field
being a monopole configuration (with minimal magnetic charge) \cite{7} on the
extra-sphere. This $U(1)$ monopole is required to support the asymptotically flat large
dimensions by cooperating with fine-tuned cosmological constant, $\lambda=2g^2/\kappa^4$.
 
From Einstein equations and the leading contribution of matters (\ref{eq6a}, \ref{eq6b}),
we obtain
\begin{equation}
\frac{\ddot{b}}{b}+\frac{\dot{b}}{b}\left(\frac{\dot{b}}{b}+3\frac{\dot{a}}{a}\right)
=-\frac{1}{b^2}+\frac{1}{4}\frac{1}{b_0^2}+\frac{3b_0^4}{b^4}+\kappa^2 c T^6\,,
\label{eq11}
\end{equation}
where $b_0^2=\kappa^2/4g^2$ is the asymptotic radius of the extra space at zero temperature
and $c=(31\pi^3/945)N_F$. In (\ref{eq11}), we impose $|\Phi|=1$ implicitly, since the
energy-difference between the different gauge vacua is order of the next-to-leading terms
in (\ref{eq9a}, \ref{eq9b}) and is expected to be negligible. We also neglect the quantum
vacuum correction because it is expected to be $\propto 1/b^6$ and this can be negligible if
$b_0^2/e^2\gg 1$. We will comment on it later.

If any value for $b$ that leads to the nonvanishing right-hand side of (\ref{eq11}), there
is no equilibrium solution ($b=$const) -of the EirBtein equation. Therefore at
\begin{equation}
T< T_{comp}=\left(\frac{31\pi^3}{945}\kappa^2N_Fb_0^2\right)^{-1/6}\,,
\label{eq12}
\end{equation}
the compactification transition takes place.

Next, we consider the symmetry breaking, in other words, the shift of the vacuum
value of $\Phi$. We compare the free energy at $|\Phi|=1$ and $\Phi=0$. If the following
condition is satisfied,
\begin{equation}
\frac{1}{4e^2b_0^4}+f(\Phi=0)<f(|\Phi|=1)\,,
\label{eq13}
\end{equation}
the ``symmetric vacuum'' with $\Phi=0$ is realized in the hot universe. In the above
equation, the value of $b$ is approximated by its equilibrium value
$b_0$. The first term in (\ref{eq13}) is the ``tree'' contribution, i.e.,
$\propto\langle{\rm Tr~}({\bf F}_{mn}{\bf F}^{mn})\rangle$. Thus we obtain the critical
temperature $T_{sb}$ of the phase transition of $\Phi$ field,
\begin{equation}
T_{sb}=\left(\frac{7\pi}{90}e^2N_Fb_0^2\right)^{-1/4}
\label{eq14}
\end{equation}
When the temperature of the universe decreases below $T_{sb}$, then phase
transition associated with the gauge configurations occurs, and at the
same time, cosmic strings of the type described in Ref.~\cite{10} are
expected to form by the Kibble mechanism.\cite{19}
 
Note that the phase associated with the monopole configuration of the gauge field
exists only when the fermionic matter dominate; for bosonic matters
$f(|\Phi|=1)-f(\Phi=0)<0$ in general.

Here we compare $T_{sb}$ with $T_{comp}$. The ratio is given by
\begin{equation}
\frac{T_{sb}}{T_{comp}}=\left(\frac{5535360\pi^3}{16807}N_F^{-1}\frac{\kappa^4}{b_0^2e^6}
\right)^{1/12}
\sim \left(10^4N_F^{-1}\frac{\kappa^4}{b_0^2e^6}
\right)^{1/12}
\end{equation}
 
We estimate this ratio by the following plausibility arguments. $\kappa^4/(b_0^2e^6)$ can be
rewritten as $(8\pi G_N/b_0^2)^2\cdot(b_0^2/e^2)^3$, where $8\pi G_N\equiv\kappa^2/b_0^2$.
$G,_N$ is interpreted as the Newton constant in four-dImensions. $8\pi G_N/b_0^2$ gives the
magnitude of the square of the coupling constant of the ``Kaluza-Klein gauge field'', which
is produced from the off-diagonal metric and in our case the gauge group is
$SU(2)$.\cite{20} $e^2/b_0^2$ can be interpreted as four-dimensional gauge
coupling. The square of couplings can be estimated to be of order
$10^{-1}$.
 
According to the estimation, we find that $T_{sb}< T_{comp}$, is satisfied only if $N_F
>10^5$. Thus, unless there exist an enormous number of fermionic degrees of freedom, the
formation of the cosmic strings must occur at the same time as the compactification
by local fluctuations of the gauge field.
 
We may consider more complicated models with larger symmetry groups. It is
known that various choices of gauge groups and internal spaces lead to many kinds
of gauge-Higgs systems.\cite{21} The high-temperatnre behavior of the models and the
possible topological defects must be thoroughly investigated.
 
In this paper we examine a hot universe scenario. Of course one may
consider a cold-universe scenario; for example, quantum fluctuation in a
curved space-time drives the compactification as well as the vacuum value
of fIelds\cite{22} In such a case, it is expected that the transitions of
compactification and the string fomations take place at the same
time.

In the above analysis, quantum correction at zero-temperature was
neglected. The corrections may always be neither well-defined nor
regularization-independent in the higher-dimensional theories.\cite{23} A detailed
investigation of this point including the quanturn corrections will be reported elsewhere.
We are also interested in the study of the production rate and cosmological application of
the cosmic strings in our model.

The width of strings in our model is the same as the radius of the
compactified space. If this scale is of order of Planck scale, the
strings may be very harmful to the evolution of the early universe.

\section*{Acknowledgements}

The authors would like to thank H. Minakata for reading an earlier
version of this manuscript. They also thank S. Hirenzaki for some
comments. This work is supported in part by the Grant-in Aid for
Encouragement of Young Scientist from the Ministry of Education, Science
and Culture (\# 6379015O). One of the authors (K. S. ) is grateful to the
Japan Society for the Promotion of Science for the fellowship. He
also thanks Iwanami F\=ujukai for financial aid.



\begin{thebibliography}{99}
\bibitem{1} Th. Kaluza, Sitz. Preuss. Akad. Wiss. Berlin Math. Phys.
{\bf K1} (1921) 966.

O. Klein, Z. Phys. {\bf 37} (1926) 895.
\bibitem{2} T. Appelquist, A. Chodos and P. G. O. Freund, Modern
Kaluza Klein Theories (Benjamin-Cummings, New York, 1987).

D. Bailin and A. Love, Rep. Prog. Phys. {\bf 50} (1987) 1087.
\bibitem{3} M. B. Green, J. H. Schwarz and E. Witten, Superstring
Theory, two volumes (Cambridge University Press, 1987).
\bibitem{4} Supergravities in Diverse Dimensions, two volumes, ed.
A. Salam and E. Sezgin (World Scientific Pub., 1989).
\bibitem{5} E. Witten, in The Proceeding of Second Shelter Island
Meeting 1983, ed. R. Jackiw et al. (MIT Press, 1985).
\bibitem{6} D. J. Gross, J. A. Harvey, E. Martinec and R. Rohm, Nucl.
Phys. {\bf B256} (1985) 253; ibid. B267 (1986) 75.
\bibitem{7} S. Randjbar-Daemi, A. Salam and J. Strathdee, Nucl.
Phys. {\bf B214} (1983) 491.
\bibitem{8} Y. Hosotani, Phys. Lett. {\bf B126} (1983) 309.

K. Shiraishi, Z. Phys. {\bf C35} (1987) 37; Prog. Theor. Phys. {\bf 80}
(1988) 601; Can. J. Phys. {\bf 68} (1990) 357.
\bibitem{9} Y. Hosotani, Phys. Lett. B129 (1983) 193; Phys. Rev. {\bf
D29} (1984) 731.
\bibitem{10} A. Nakamula, S. Hirenzaki and K. Shiraishi, Nucl.
Phys. {\bf B339} (1990) 533.
\bibitem{11} H. B. Nielsen and P. Olesen, Nucl. Phys. {\bf B61} (1973)
45.
\bibitem{12} R. Rajaraman, Solitons and Instantons (North-Holland,
Amsterdam, 1982).
\bibitem{13} T. Eguchi, P. B. Gilkey and A. J. Hanson, Phys. Rep. {\bf
66} (1980) 213.
\bibitem{14} K. Shiraishi and A. Nakamula, Mod. Phys. Lett. {\bf A5}
(1990) 1109.
\bibitem{15} S. Coleman and E. Weinberg, Phys. Rev. {\bf D7} (1973)
1888.

R. H. Brandenberger, Rev. Mod. Phys. {\bf 57} (1985) 1.
\bibitem{16}  F. S. Accetta and E. W. Kolb, Phys. Rev. {\bf D34} (1986)
1798.
\bibitem{17} I. G. Moss, Phys. Lett. {\bf B140} (1984) 29.
\bibitem{18} J. S. Dowker, Phys. Rev. {\bf D29} (1984) 2773; Class.
Q. Grav. {\bf 1} (1984) 359.

See also A. Nakamula and K. Shiraishi, Nuovo Cim. {\bf B105} (1990) 179.
\bibitem{19} T. W. B. Kibble, J. of Phys. {\bf A9} (1976) 1387.

A. Vilenkin, Phys. Rep. {\bf 121} (1985), 263.
\bibitem{20} S. Weinberg, Phys. Lett. {\bf B125} (1983) 265.
\bibitem{21} Yu. A. Kubyshin, J. M. Mourao and I. P. Volobujev,
Nucl. Phys. {\bf B322} (1989) 531; Int. J. Mod. Phys. {\bf A4} (1989)
151.
\bibitem{22} A. D. Linde, Phys. Lett. {\bf B116} (1982) 335; ibid. {\bf B130}
(1983) 330.

A. Vilenkin and L. Ford, Phys. Rev. {\bf D26} (1982) 1231.

A. A. Starobinsky, in Current Topics in Field Theory, Quantum Gravity
and Strings, ed. H. J. de Vega and N. Sanchez (Springer, Berlin, 1986).
\end{thebibliography}
\end{document}